\documentclass[11pt,a4paper]{article}

\usepackage[utf8]{inputenc}
\usepackage{mathrsfs,amsmath} 
\usepackage{systeme}
\usepackage{mathtools}
\usepackage{authoraftertitle}
\usepackage{physics}
\usepackage{hyperref}
\usepackage{subcaption} 
\usepackage{lipsum}
\usepackage[draft]{todonotes} 
\usepackage{chngcntr}
\usepackage{tikz-cd}
\usepackage{chemfig}
\usepackage{needspace}
\usepackage{url}
\usepackage{cleveref}
\usepackage[absolute]{textpos}

\usepackage{mathrsfs,amsmath} 
\usepackage[section]{placeins}
\usepackage{subfiles}

\usepackage[math]{cellspace}

\usepackage{graphicx,epsf}
\usepackage{array,multirow}
\usepackage{bm}

\renewcommand{\vec}[1]{\boldsymbol{#1}}

\newcommand{\ySubFig}[1]{
	\begin{subfigure}[t]
		{0.49\textwidth}
		\includegraphics[width=\textwidth]{./#1}
	\end{subfigure}
}

\newcommand{\yFigTwo}[4]{
	\begin{figure}[!ht]
		\ySubFig{#1}\ySubFig{#2}\caption{#3 \label{#4}}
	\end{figure}
}

\newcommand{\quotes}[1]{``#1''}

\begin{document}

\begin{center}
	{\large{\textbf{Gyrokinetic investigation of the nonlinear interaction of Alfv\'en 
instabilities and energetic-particle driven geodesic acoustic modes}}}\\
	\vspace{0.2 cm}
	\author{F. Vannini}
	{\underline{F. Vannini}$^{1}$, A. Biancalani$^1$, A. Bottino$^1$, T. Hayward-Schneider$^{1}$, Ph. Lauber$^1$, A. Mishchenko$^2$, E. Poli$^1$, G. Vlad$^3$ and the ASDEX Upgrade team$^4$.}
	
	\vspace{0.2 cm}

{
$^1$Max-Planck-Institut f\"ur Plasmaphysik, 85748 Garching, Germany \\
$^2$Max-Planck-Institut f\"ur Plasmaphysik, 17491 Greifswald, Germany\\
$^3$ENEA, Fusion and Nuclear Safety Department, 00044 Frascati (Roma), Italy\\
$^4$ See author list of \textit{H. Meyer et al. 2019 Nucl. Fusion \textbf{59} 112014}\\
}
	\vspace{0.2 cm}
	\small{francesco.vannini@ ipp.mpg.de\\
		}
\end{center}

\begin{abstract}
This paper presents a study of the interaction between Alfv\'en modes and zonal structures, considering a realistic ASDEX Upgrade equilibrium. The results of gyrokinetic simulations with the global, electromagnetic, particle-in-cell code ORB5 are presented, where the modes are driven unstable by energetic particles with a bump-on-tail equilibrium distribution function, with radial density gradient. Two 
regimes have been observed: at low energetic particles concentration, the Alfv\'en mode saturates at much higher level in presence of zonal structures; on the other hand at high energetic particles concentration the difference is less pronounced. The former regime is characterized by 
the zonal structure (identified as an energetic particle driven geodesic acoustic mode), being more unstable than the Alfvén mode. In the latter regime the Alfvén mode is more unstable than the zonal structure. 
The theoretical explanation is given in terms of a 3-wave coupling of 
the energetic particle driven geodesic acoustic mode and Alfv\'en mode, mediated by the curvature-pressure coupling term of the 
energetic particles.
\end{abstract}

\newpage

\section{Introduction}
The next generation of fusion relevant machines (ITER \cite{Tomabechi_1991,ITER}, DEMO \cite{DEMO}) will be characterized by a large population of energetic particles (EPs). These are fusion products and charged particles generated by external power sources (like neutral beam injection, NBI). EPs are characterized by velocities much higher than the thermal velocity of the particles of the background plasma, $v_{EP}\gg v_{th,i,e}$, where the subscripts $i$ and $e$ refer respectively to the main ion population and to the electrons present in the plasma. The transport and confinement of the EPs in Tokamaks and  the physics related to EPs is of primary importance in order to achieve self heating plasma. 
In fact EPs can drive unstable, via wave-particle interactions, symmetry breaking electromagnetic perturbations whose presence can redistribute the EPs population, expelling them out of the plasma before they can thermalize \cite{Chen_Zonca_2016}. This consequently can lead to a less effective heating. In addition, the violent migration of the EPs toward the walls, caused by generated short but intense perturbations dubbed abrupt large-amplitude events \cite{Bierwage2018SimulationsTA}, can possibly damage the machine. The understanding of the intensity of the fields of the saturated induced instabilities together with the transport of the EPs represent a key topic to reach fusion.

In fusion devices the EPs have a characteristic periodic motion (bounce frequency, transit frequency,...) of the same order of magnitude as that typical of the shear Alfv\'en waves %
(SAWs \cite{1942Natur.150..405A}). When the drive provided by the EPs exceeds the damping of these plasma fluctuations, a broad spectrum of  Alfv\'en waves can be excited \cite{Rosenbluth_Rutherford_1975,Todo_2018}. These driven instabilities are classified into two types: Alfv\'en eigenmodes (AMs) and energetic particle modes (EPMs). The AMs are characterized by frequencies located inside the frequency gaps of the SAWs continuum spectrum, created by the field geometry and by plasma non-uniformities \cite{CHENG198521,Heidbrink_1993}. The EPMs  are non-normal modes of the SAWs continuum spectrum, emerging as discrete fluctuations at the frequency where wave-EP power is maximized \cite{Chen_1994}. The linear and nonlinear dynamics of SAWs driven instabilities has been reviewed in Ref.\cite{Chen_Zonca_2016}.
 
Additionally, EPs can drive unstable modes with frequency comparable with that of the geodesic acoustic modes (GAM, \cite{Winsor_1968}), characterized by $(m,n)=(0,0)$ scalar potential and $(m,n)=(1,0)$ up-down anti-symmetric density perturbation (being $m$ and $n$  respectively the poloidal and toroidal mode numbers). These driven modes are the energetic particle geodesic acoustic mode (EGAM), excited via free energy associated with velocity space gradients in the EPs distribution, as it has been shown analytically in Ref.\cite{Fu_PRL_2008}. Studies on the nonlinear dynamics of EGAMs have been recently presented in \cite{biancalani_egam,Qiu_2011}.

The comprehension of the dynamics of the great zoology of modes present in plasma, their interaction and the redistribution of EPs, is crucial to understand the properties of burning plasma. The study of the saturation levels of these instabilities is fundamental to be able to be predictive regarding the intensity of the fields that will be present in future reactors. This motivate the interest in the study of the nonlinear evolution of these instabilities. Traditionally, the two main saturation mechanisms of such instabilities, have been recognized as A) reduction of the drive due to the redistribution of the EP population \cite{Berk_Breizmann_1990}; B) the transfer of energy to other modes via mode-mode coupling mediated by the thermal plasma nonlinearities \cite{Zonca_Romanelli_1995}. More recently, a new mechanism has also been studied, consisting in mode-mode coupling mediated by the EP nonlinearity. This mechanism has been found to be responsible, for example, of the excitation of zonal structures by Alfv\'en modes, and it has been called forced-driven excitation \cite{Qiu_2016,Todo_2010}.

In the present paper, the interaction between AMs and EGAMs is studied showing the results of numerical simulations obtained with the nonlinear, gyrokinetic, electromagnetic, PIC code ORB5 \cite{LANTI2020107072}. The dynamics that EGAM $(m,n)=(0,0)$ and dominant AM $(m,n)=(2,1)$ exhibit in numerical simulations where only one toroidal mode number is retained ($n=\{i\}\,,\,i=0,1$) is compared  with the dynamics observed in simulations where both toroidal mode numbers are kept. The nonlinearities have been maintained only in the EPs dynamics, while the background particle species (Deuterium and electrons) follow their unperturbed trajectories. The EPs have a double-bump-on-tail distribution function in velocity space with radial density gradient. The background plasma species have Maxwellian distribution functions. 

The paper is structured as follows. In Sec.\ref{Sec:model}, the model of ORB5 is presented. In Sec.\ref{Sec:equilibrium} the equilibrium in use in the simulations is shown, together with the simulations details.
In Sec.\ref{Sec:circular_equilibrium} we investigate the coupling of AMs and EGAMs in a simplified configuration (circular flux surfaces) to simplify the physics and more easily compare with analytical theory. In Sec.\ref{Sec:wave_wave-interaction}, the analytical interpretation is provided. In Sec.\ref{Sec:NLED-AUG_case}, the application to a more realistic configuration (ASDEX Upgrade experimental magnetic equilibrium) is described. Here the results of numerical simulations  obtained with a realistic scenario, the so-called NLED-AUG case \cite{Lauber,Lauber2}, are presented. This section represents an extension of the studies detailed in Ref.\cite{Di_Siena_2018,NOVIKAU2020} and in Ref.\cite{VANNINI2020}. There, the dynamics, respectively, of EGAMs and AMs  has been individually investigated, retaining only one toroidal mode number. In this work the main novelty is that, in order to study the interaction of the EGAM, with the dominant Alfv\'en mode (AM), $(m,n)=(2,1)$ (which is identified as an EPM), both the toroidal modes $n=\{0,1\}$ are retained in the performed simulations.

\newpage
\section{The Model}\label{Sec:model}
ORB5 \cite{LANTI2020107072} is a nonlinear, global, electromagnetic, particle in cell (PIC) code which solves the gyrokinetic Vlasov-Maxwell system of equations \cite{Tronko_2016,Tronko_2017}, accounting for the presence of collisions and sources.  The code uses a system of straight field-line coordinates, $(s,\theta^{\ast},\varphi)$. The poloidal flux $\psi$, normalized at its value at the edge $\psi_{0}$, plays the role of radial coordinate ($s=\sqrt{\psi/\psi_{0}}$, $0 \leq s \leq 1$). $\varphi$ is the toroidal angle while the poloidal magnetic $\theta^{*}$ angle is defined as:
\begin{equation}
    \theta^{\ast}=\frac{1}{q(s)}\int_{0}^{\theta}\frac{\Vec{B}\cdot\nabla\phi}{\Vec{B}\cdot\nabla\theta^{\prime}}d\theta^{\prime}
\end{equation}
being $q(s)$ the safety factor profile, $\theta^{\prime}$ the geometric poloidal angle and $\Vec{B}$ the background magnetic field, linked to the magnetic potential $\Vec{A}_{0}$ through the equation $\Vec{B}=\nabla\cross \Vec{A}_{0}$ 

In ORB5 all the physical quantities are normalized to four reference parameters: the mass and charge of the main ion species ($m_{i}$ and $q_{i}=e Z_{i}$, being $e$ the elementary charge and $Z_{i}$ the atomic number of the $i$-th ion species), the values of the electron temperature at a radial location $s_{0}$ and of the magnetic field on-axis ($T_{e}(s_{0})$ and $B_{0}$ respectively). The derived units are obtained from these four parameters. For example, the time is given in units of the inverse of the ion cyclotron frequency $\omega_{ci}=q_{i}B_{0}/(m_{i}c)$ (being $c$ the speed of light in vacuum), the velocities are multiple of the ion sound velocity $c_{s}=\sqrt{T_{e}(s_{0})/m_{i}}$, 
 and so on. 
 
 In ORB5 is possible to consider both analytical equilibrium, comprising circular magnetic surfaces, and ideal-MHD equilibria. This is a solution of the Grad-Shafranov equation, calculated through the CHEASE code \cite{CHEASE}. 
 
 The species distribution function $f_{s}$ is divided into a prescribed time-independent background distribution function $F_{0,s}$ and into a time dependent part $\delta f_{s}$. The subscript $s$ refers to the particle species (that is $s=i,e,f$, respectively background ions, electrons and fast particles). The time dependent part of the distribution function is sampled with numerical particles, called markers, representing a portion of the phase space. The gyrokinetic Vlasov equation for the perturbed distribution function is:
\begin{equation}
    \frac{d }{dt}\delta f_{s}=-\dot{\Vec{R}}\cdot\frac{\partial F_{0,s}}{\partial \Vec{R}}\bigg\rvert_{\mathcal{E},\,v_{\parallel}} - \dot{\mathcal{E}}\frac{\partial F_{0,s}}{\partial \mathcal{E}}\bigg\rvert_{\vec{R},\,v_{\parallel}}
    - \dot{v}_{\parallel}\frac{\partial F_{0,s}}{\partial v_{\parallel}}\bigg\rvert_{\vec{R},\,\mathcal{E}}
    \quad \mathcal{E}=\frac{v_{\parallel}^{2}}{2}+\mu B \quad \mu=\frac{v_{\perp}^{2}}{2 B}
    \label{Eq:delta_f}
\end{equation}
for the bulk ions, $F_{0}(\vec{R},\mathcal{E})$ only, while for a bump-on-tail $F_{0}(\vec{R},\mathcal{E},v_{\parallel})$. Details can be found in Ref.\cite{Zarzoso_2014}. In Eq.\ref{Eq:delta_f} $\Vec{R}$ is the gyrocenter trajectories, $v_{\parallel}$ and $v_{\perp}$ are the parallel and perpendicular velocities respect to the background magnetic field. The equations of motions in mixed-variable formulation of the gyrocenter characteristics $(\Vec{R},\mathcal{E},v_{\parallel},\mu)$ are \cite{ami_2019}:
\begin{multline}
    \Vec{\dot{R}} =v_{\parallel}\Vec{\hat{b}} -v_{\parallel}^{2}\frac{c\,m_{s} }{q_{s}B_{\parallel}^{*}}\Vec{\hat{b}}\cross\left(\Vec{\hat{b}}\cross\curl{\Vec{b}}\right)+\mu\frac{c\,m_{s}}{q_{s}B_{\parallel}^{*}}\Vec{\hat{b}}\cross\nabla B+\epsilon_{\delta}\left[
    \frac{\hat{\Vec{b}}}{B_{\parallel}^{*}}\cross\nabla \langle \phi-v_{\parallel}A_{\parallel}^{h}-v_{\parallel}A_{\parallel}^{s}\rangle-\frac{q_{s}}{m_{s}}\langle A_{\parallel}^{h}\rangle\hat{\Vec{b}}^{*}\right]
    \label{Eq:ORB1}
\end{multline}
\begin{multline}
    \dot{v}_{\parallel}=\mu B\nabla\cdot \Vec{b}+\mu v_{\parallel}\frac{c\,m_{s}}{q_{s}B_{\parallel}^{*}}\left(\Vec{\hat{b}}\cross \left(\Vec{\hat{b}}\cross\curl{\Vec{b}}\right)\right)\cdot\nabla B+\\
    -\epsilon_{\delta}\bigg\{\mu\frac{\Vec{\hat{b}}\cross \nabla B}{B_{\parallel}^{*}}\cdot \nabla\langle A_{\parallel}^{s} \rangle
    +\frac{q_{s}}{m_{s}}\left[\hat{\Vec{b}}^{*}\cdot\nabla\langle \phi -v_{\parallel}A_{\parallel}^{h}\rangle+\frac{\partial }{\partial t}\langle A_{\parallel}^{s}\rangle  \right]\bigg\}
    \label{Eq:ORB2}
\end{multline}
\begin{equation}
    \dot{\mathcal{E}}=v_{\parallel}\dot{v}_{\parallel}+\mu\nabla B\cdot \dot{\Vec{R}}
     \label{Eq:ORB3}
\end{equation}
\begin{equation}
    \dot{\mu}=0
    \label{Eq:ORB4}
\end{equation}
where the terms proportional to $\epsilon_{\delta}$ are the nonlinear terms, corresponding to the perturbed equations of motion and $\langle ...\rangle$ is the gyro-average. \Cref{Eq:ORB1,Eq:ORB2,Eq:ORB3} are formulated in mixed-variables formulation \cite{ami_2014}. Here, the perturbed magnetic potential $A_{\parallel}$ has been split into its symplectic and hamiltonian parts: $A_{\parallel}=A^{s}_{\parallel} + A^{h}_{\parallel}$. In \Cref{Eq:ORB1,Eq:ORB2,Eq:ORB3} the modified vector potential is present $\Vec{A}^{*}$, the modified parallel magnetic filed $B_{\parallel}^{*}$ and the unit vector in the direction of the magnetic filed, are present:
\begin{equation}
    \Vec{A}^{*}=\Vec{A}+(m_{s}v_{\parallel}/q_{s})\hat{\Vec{b}} \quad \Vec{B}^{*}=\nabla\cross \Vec{A}^{*}\quad \Vec{\hat{b}}=\Vec{B}/B
\end{equation} 

The equations of motion are coupled with the following field equations: the quasineutrality condition, the parallel Ampère's law and the ideal Ohm's law:

\begin{equation}
-\nabla\cdot \left[\left(\sum_{s=i,f}\frac{q_{s}^{2}n_{s}}{T_{s}}\rho_{s}^{2}\right)\nabla_{\perp}\phi\right]=\sum_{i,e,f}q_{s}n_{1,s}\quad n_{1,s}= \int dW \langle \delta f_{s}\rangle   
\label{Eq:quasineutrality}
\end{equation}

\begin{equation}
    \left(\sum_{i,e,f}\frac{\beta_{s}}{\rho_{s}^{2}}-\nabla_{\perp}^{2}\right)A_{\parallel}^{h}=\mu_{0}\sum_{i,e,f}j_{\parallel,1,s}+\nabla^{2}_{\perp}A_{\parallel}^{s} \quad j_{\parallel,1,s}=q_{s}\int dW v_{\parallel}\langle \delta f_{s}\rangle
    \label{Eq:Ampere}
\end{equation}

\begin{equation}
    \frac{\partial }{\partial t}A_{\parallel}^{s}+\hat{\Vec{b}}\cdot \nabla\phi = 0
\end{equation}
where:
\begin{equation}
    n_{s}=\int dW F_{0,s}\quad \beta_{s}=\mu_{0}\frac{n_{s}T_{s}}{B_{0}^{2}}
\end{equation}
The integrals in  \cref{Eq:quasineutrality,Eq:Ampere} represent respectively the mixed-variable gyrocenter density and the mixed-variable gyrocenter current. These integrals are calculated in the phase-space volume $dW=B_{\parallel}^{*}d v_{\parallel}d\mu d\alpha$ (being $\alpha$ the gyro-phase). $\Vec{\rho}$ is the particle gyroradius, while $\rho_{s}=\sqrt{m_{s}T_{s}}/(q_{s}B)$ is the thermal gyroradius. The equations are solved through the mixed-variable pullback algorithm, presented in Ref.\cite{ami_2014}, which has been able to mitigate the so-called cancellation problem, \cite{cancellation}. 
The typical modes of interest are mainly aligned with the magnetic field line $m\simeq n q(s)$. So a filter is applied to the Fourier coefficients of the perturbed density and current \cite{Jolliet2007AGC}. In this way all the nonphysical modes introduced by charge and current deposition are filtered out. For each toroidal mode $n\in [n_{min},n_{max}]$ only the poloidal modes $m\in [-n q(s)\pm \Delta m]$, where $\Delta m$ is the width of the retained poloidal modes.

In the present paper we will study the nonlinear interaction of modes, analyzing simulations where only the EPs full dynamic is retained. This means that only in the EPs equations of motion, the nonlinear terms will be present, that is the terms proportional to $\epsilon_{\delta}$ in \cref{Eq:ORB1,Eq:ORB2}. Electrons finite Larmor radius effects are neglected, given $\rho_{e}\longrightarrow 0$.

\newpage
\section{Equilibrium}\label{Sec:equilibrium}
The discharge $\#$31213@0.84s has been chosen as a base case for linear and nonlinear EPs simulations. The uniqueness of this scenario, the so called NLED-AUG case \cite{Lauber}, is due to the fact that it exhibits a neutral beam (NB) induced fast-ion $\beta$ comparable to that of the background plasma. In addition, the fast ions have energy 100 times larger than the thermal background. 
These unexplored corner of plasma parameters has been chosen to match the realistic ratios of plasma parameters that are going to be met in future fusion machines and to obtain a scenario where the transport of fast particles and the induced mode dynamics can be mainly attributed to the presence of the EPs being the effects of the background plasma minimized \cite{Lauber2}. This scenario is rich of nonlinear physics. A TAE burst is observed to trigger EGAMs, suggesting the coupling of these modes via the velocity space (EPs avalanches) and via mode-mode coupling processes. The great variety of nonlinear physics present here, together with the fact that the mode dynamics is mainly mediated by the EPs, makes of this an important scenario for the validation of theoretical tools and codes.

\begin{figure}[!h]
     \centering
	\includegraphics[width=0.5\textwidth]{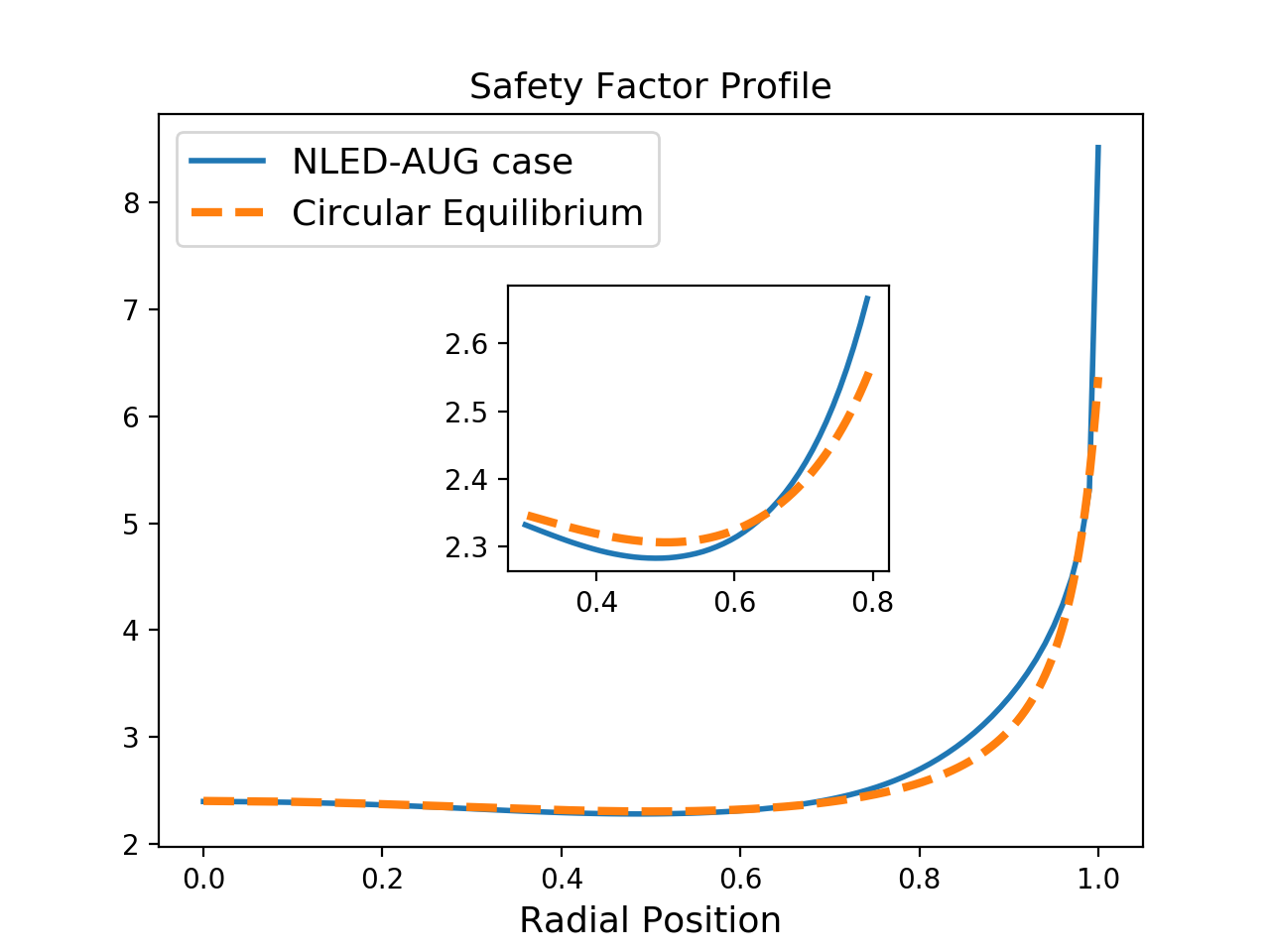}
	\caption{Radial dependence of the safety factor profile of the NLED-AUG case and of circular equilibrium (dashed line) discussed in the  Sec.\ref{Sec:equilibrium}.}
	\label{Fig:q_profile}
\end{figure}

The safety factor profile has a reversed shear, with a minimum located at the radial position $s\simeq 0.5$, in the amount of $2.2$. In Fig.\ref{Fig:q_profile} the $q$-profile of the NLED-AUG case is shown (blue curve). The background plasma temperature profiles and the electron density profile of the NLED-AUG case are shown in  Fig.\ref{Fig:profiles_radial}.

\yFigTwo
{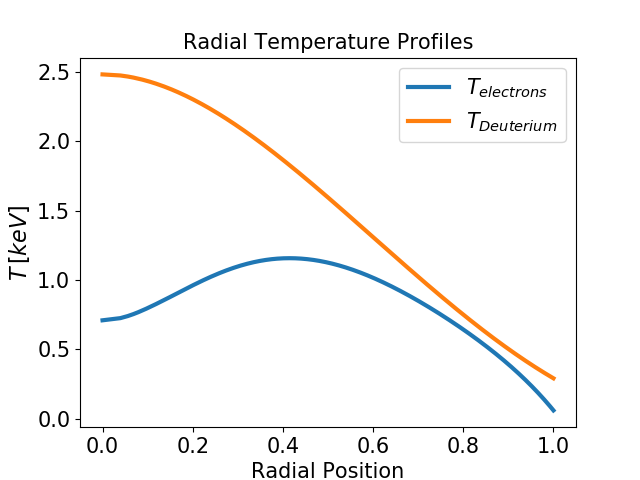}
{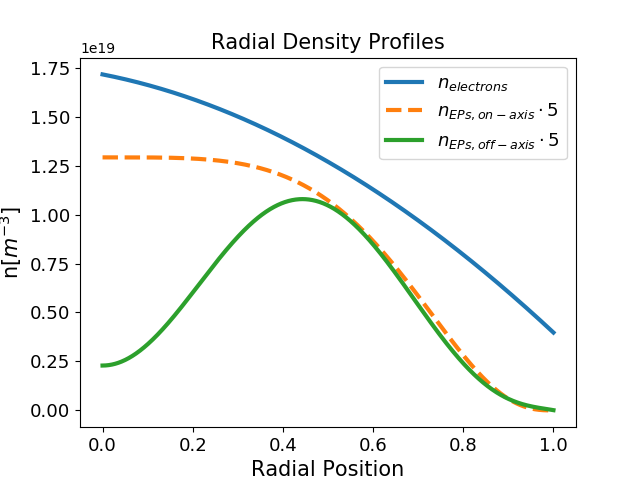}
{Radial dependence of the temperature and density profiles. On the right, the radial density profiles of the EPs are shown, for the NLED-AUG case (green line) and for the present case (orange dotted line).}
{Fig:profiles_radial}

In Tab.\ref{tab:1}, the values of some important constants used in the simulations are shown. 
\begin{table}[htbp]
	\centering
	\vspace{1ex}	
	\begin{tabular}{||c|c|c|c|c|c||}\hline
		
		\hline
	     $a_{0}\,[m]$&$R_{0}\,[m]$&$B_{0}\,[T]$&$\omega_{ci}\,[rad/s]$&$\beta$&$L_{x}$\\
	     \hline
	     $0.482$&$1.666$&$2.202$&$1.055\cdot 10^{8}$&$2.7\cdot 10^{-4}$&$551.6$\\
		\hline
	\end{tabular}
	\caption{Constants in use: averaged minor radius, major radius, magnetic field on axis, ion cyclotron frequency, normalized plasma pressure. $L_{x}$ is the normalized size of the plasma system, defined as $L_{x}=2\,a_{0}\frac{\omega_{ci}}{c_{s}}$ , being $c_{s}$ the ion sound velocity.}
	\label{tab:1}
\end{table}

In Sec.\ref{Sec:circular_equilibrium} and Sec.\ref{Sec:NLED-AUG_case}, we will show the results of simulations obtained with the radial electron density profile and temperature profiles of the NLED-AUG case. The main ion species and the EPs are deuterium plasma. The equilibrium quasi neutrality is fulfilled by keeping constant the radial electron density profile and varying the EPs concentration, together with the deuterium concentration, satisfying: $n_{e}=\sum_{i}Z_{i}n_{i}$ (being $n$ the density profile of the $i$-th species). 
The bulk plasma species (electrons and deuterium) have Maxwellian distribution functions, while the EPs have a double bump-on-tail distribution function (as in \cite{Di_Siena_2018},\cite{NOVIKAU2020},\cite{Zarzoso_2014},\cite{NOVIKAU2019}), because an anisotropy in velocity is needed to drive unstable EGAMs, \cite{Qiu_2011}:

\begin{equation}
    f_{EPs}\sim \frac{1}{2}\left[ e^{-\frac{(\mathcal{E}-v_{\parallel}v_{\parallel,0}+v_{\parallel,0}^{2}/2)}{tval}}+ e^{-\frac{(\mathcal{E}+v_{\parallel}v_{\parallel,0}+v_{\parallel,0}^{2}/2)}{tval}}\right]
    \label{Eq:bouble-bump-on-tail}
\end{equation}
The local maximum of the distribution function, in the velocity space, is located at $v_{\parallel}=v_{\parallel,0}$, while $tval$ is the width in velocity space of the two shifted Maxwellians.

In Sec.\ref{Sec:circular_equilibrium}, a circular magnetic equilibrium will be considered and will be characterized by a safety factor profile having a radial dependence close to that of the NLED-AUG case (see Fig.\ref{Fig:q_profile}, orange dotted line). The EPs will have an on-axis radial density profile (see Fig.\ref{Fig:profiles_radial} on the right, the orange dotted curve).
Through this approximation, we have that the EPs provide a contribution with constant sign to the linear growth rate of the driven modes, through the spatial derivative of their distribution function \cite{Todo_2018,VANNINI2020,Betti_1992}, being:
\begin{equation}
    \gamma^{L}_{i}\sim \omega \frac{\partial f_{i}}{\partial \mathcal{E}}-\frac{n}{q_{i}}\frac{\partial f_{i}}{\partial \psi}
\label{Eq:linear_growth_rate}
\end{equation}
In Eq.\ref{Eq:linear_growth_rate}, the poloidal flux $\psi$ plays the role of radial coordinate, $f_{i}$ and $q_{i}$ are, respectively, the distribution function and charge of the $i-$th species of the plasma.

In Sec.\ref{Sec:NLED-AUG_case} we will adopt the realistic ASDEX Upgrade magnetic equilibrium. The EPs will have an off-axis radial density profile (see Fig.\ref{Fig:profiles_radial} on the right, the green curve), as modelled by TRANSP \cite{TRANSP}.

\clearpage

\newpage

\section{Modification of the AM saturation in the presence of ZS}\label{Sec:Results}

\subsection{Basic physics of the AM/ZS interaction}\label{Sec:circular_equilibrium}

In this section, we focus on the basic physics of the interaction between AMs and Zonal Structures (ZS) indicating, with this term, axysimmetric perturbations in general, that can be: Zero Frequency Zonal Flows (ZFZF), geodesic acoustic modes (GAM) or energetic particle driven geodesic acoustic modes (EGAM). To this aim, we consider here a simplified configuration, where the flux surfaces are circular and concentric. This allows us to neglect the secondary correction due to the geometry. 

For computational reasons, in Sec.\ref{Sec:circular_equilibrium} the electron mass has been then taken 500 times lighter than the ions mass $m_{e}=m_{i}/500$. In Tab.\ref{tab:2}, the values of some important parameters used in the simulations are shown. 
\begin{table}[htbp]
	\centering
	\vspace{1ex}	
	\begin{tabular}{||c|c|c|c|c||}\hline
		
		\hline
	     $nptot_{D,e,EP}\cdot 10^{7}$&$\Delta t\,[\Omega_{ci}^{-1}]$&$n_{s}$&$n_{\theta^{*}}$&$n_{\varphi}$\\
	     \hline
	     $3,12,3$&$3$&$288$&$288$&$48$\\
		\hline
	\end{tabular}
	\caption{Main simulations parameters (number of markers, time step, grid points).}
	\label{tab:2}
\end{table}

In Fig.\ref{Fig:evolution_modes_ad_hoc} we show the simulation results obtained in two regimes: at low ($\langle n_{EP}\rangle /\langle n_{e}\rangle =0.0379$) and high ($\langle n_{EP}\rangle/\langle n_{e}\rangle=0.114$) EPs concentration (being $\langle..\rangle$ the volume average). For each regime, we present the mode dynamics observed in simulations where only a single toroidal mode was retained, that is only $n=\{0\}$ or $n=\{1\}$. These correspond respectively to the green and red curves in Fig.\ref{Fig:evolution_modes_ad_hoc}. We compare this evolution with the dynamics observed in simulations where both toroidal modes are present $n=\{0,1\}$ (blue and orange curves in Fig.\ref{Fig:evolution_modes_ad_hoc}). 

The measured growth rates and frequencies will be provided in units of Alfv\'en frequencies measured on axis: $\omega_{A0}=\frac{1}{R_{0}}\sqrt{\frac{B_{0}}{4\pi\rho_{m,0}}}$, being $\rho_{m,0}$ the value of the background plasma density on axis: $\rho_{m,0}=m_{i}n_{i}+m_{e}n_{e}$.

In the regime at low EPs concentration, we observe the ZS (green curve in Fig.\ref{Fig:evolution_modes_ad_hoc}) to be more unstable than the dominant AM (red curve in Fig.\ref{Fig:evolution_modes_ad_hoc}). Their mode structure is shown in Fig.\ref{Fig:mode_structure_ad_hoc_single_mode} and the measured growth rates and frequencies are reported in Tab.\ref{tab:gro_freq_adhoc}. The ZS is identified as an EGAM while the AM (mainly peaked around $s\simeq 0.5$) has a frequency sitting slightly below the SAW continuum branch $(m,n)=(2,1)$. 

\begin{table}[htbp]
	\centering
	\vspace{1ex}	
	\begin{tabular}{||c|c|c||}\hline
		\hline
	     &$\gamma^{L}\,[\omega_{A0}]$&$\omega\,[\omega_{A0}]$\\
	     \hline
	     $AM$&$(2.43\pm0.05)\cdot 10^{-3}$&$-0.0985\pm 0.0005$\\
	     \hline 
	     $EGAM$&$(5.4\pm0.1)\cdot 10^{-3}$&$0.063\pm0.001$\\
		\hline
	\end{tabular}
	\caption{Growth rates and frequencies of the dominant modes in the simulations where only one toroidal mode number is retained. Low EPs concentration.}
	\label{tab:gro_freq_adhoc}
\end{table}
In the simulation where both the toroidal modes are present, we observe the EGAM dynamics (blue curve in Fig.\ref{Fig:evolution_modes_ad_hoc}) to be practically unaffected by the presence of the AM in the linear phase. On the contrary, the AM dynamics (orange curve in Fig.\ref{Fig:evolution_modes_ad_hoc}) appears to be driven by the EGAM. This results in an increase of the AM drive and of its frequency, lying now on the Alfv\'en continuum. These values have been measured in the temporal domain $t[\omega_{ci}^{-1}]\in[28000;39000]$:
\begin{equation}
    \gamma^{NL}_{AM}=(7.6\pm 0.4)\cdot 10^{-3}\omega_{A0}\quad \omega^{NL}_{AM}=-0.14\pm 0.1\,\omega_{A0}
\end{equation}
These values of growth rate and frequency have been measured in the linear phase of the mode dynamics. They have been labelled with the superscript $NL$ (standing for \textit{nonlinear}), to underline that these values arise from the nonlinear interaction between the AM and the EGAM, as opposed to what happened in the simulations where a single toroidal mode was present. Another qualitative proof of the mode-mode interaction is provided by Fig.\ref{Fig:mode_structure_ad_hoc}. There, we show the observed mode structure for each toroidal mode number in simulations where both are present $n=\{0,1\}$. We show the mode structure at a time before the saturation level is reached (around $t\sim 40000\,\omega_{ci}^{-1}$). In Fig.\ref{Fig:mode_structure_ad_hoc} we observe the peaks of the electrostatic potential of $(m,n)=(2,1)$, located near the nodes of the electrostatic potential $(m,n)=(0,0)$. This suggests that the energy is going from the EGAM to the AM and the acting interaction mechanism is mode-mode coupling, as will be explained below.

In the regime at higher EP concentrations (Fig.\ref{Fig:evolution_modes_ad_hoc}) we observe the AM to be more unstable than the ZS:
\begin{equation}
    \gamma^{L}_{AM}=(1.078\pm 0.002)\cdot 10^{-2}\omega_{A0}\quad \gamma^{L}_{ZS}=(6.3\pm 0.1)\cdot 10^{-3}\omega_{A0}\,\,.
\end{equation}
When the two modes are together, we observe the AM dynamics to be practically unchanged, instead the mode $(m,n)=(0,0)$, that here we identify as a ZFZF, has a growth rate:
\begin{equation}
    \gamma^{NL}_{ZS}=(1.9\pm 0.5)\cdot 10^{-2}\omega_{A0}\,\,.
\end{equation} 
The AM pumps the ZS, with a forced-driven mechanism,  that was analytically derived in Ref.\cite{Qiu_2016}. In Ref.\cite{Qiu_2016} a pumping TAE with its complex conjugate was found to be responsible for the drive of the ZFZF, growing with: $\gamma^{NL}_{ZFZF}=2\gamma_{AM}$. 

In Sec.\ref{Sec:wave_wave-interaction}, we extend the analytical calculation of Ref.\cite{Qiu_2016}, to show that a general AM can force-drive a ZS and to explain the inverse mechanism, namely the excitation of an AM by a ZS, which is what is observed in the regime at low EPs concentration.

\yFigTwo
{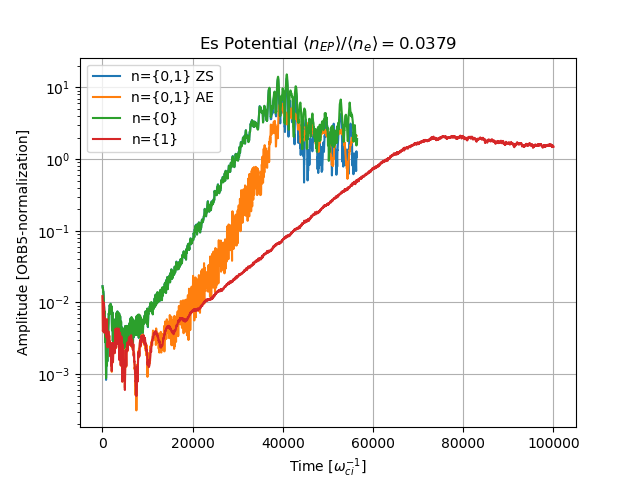}
{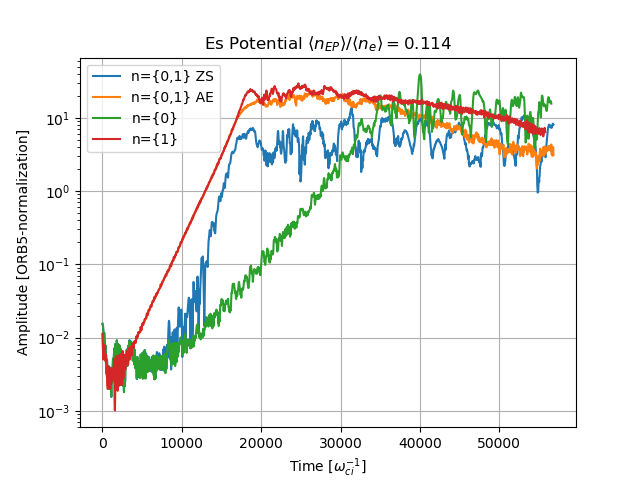}
{Simulations at low (left) and high (right) EPs concentration.}
{Fig:evolution_modes_ad_hoc}

\yFigTwo
{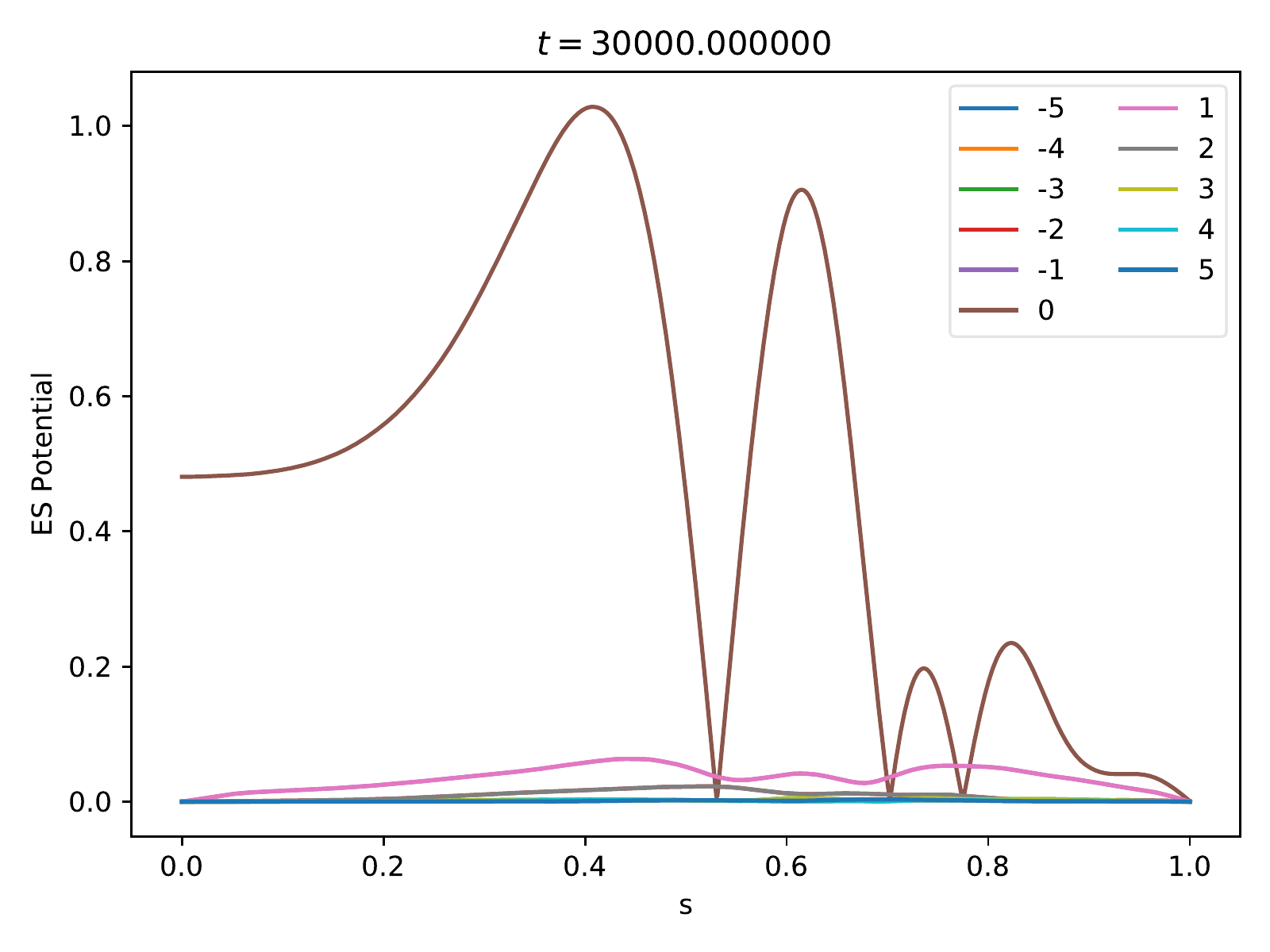}
{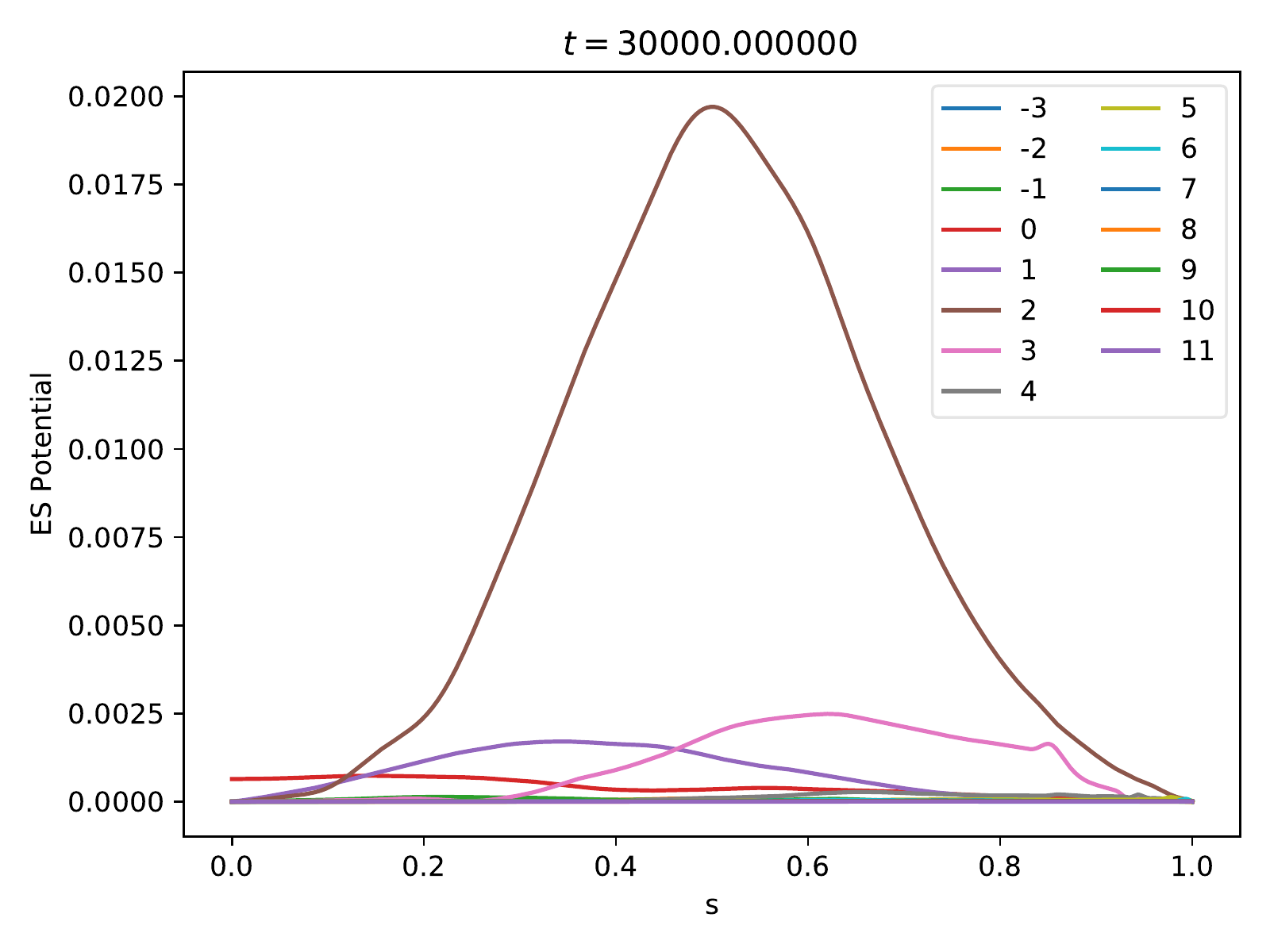}
{Radial dependence of the absolute value of $n=\{0\}$ 
(left) and $n=\{1\}$ (right) scalar potential components of the simulation at low EPs concentration (see Fig.\ref{Fig:evolution_modes_ad_hoc} on the left). Here only one toroidal mode number is retained. Color label correspond to different poloidal harmonics $m$.}
{Fig:mode_structure_ad_hoc_single_mode}

\yFigTwo
{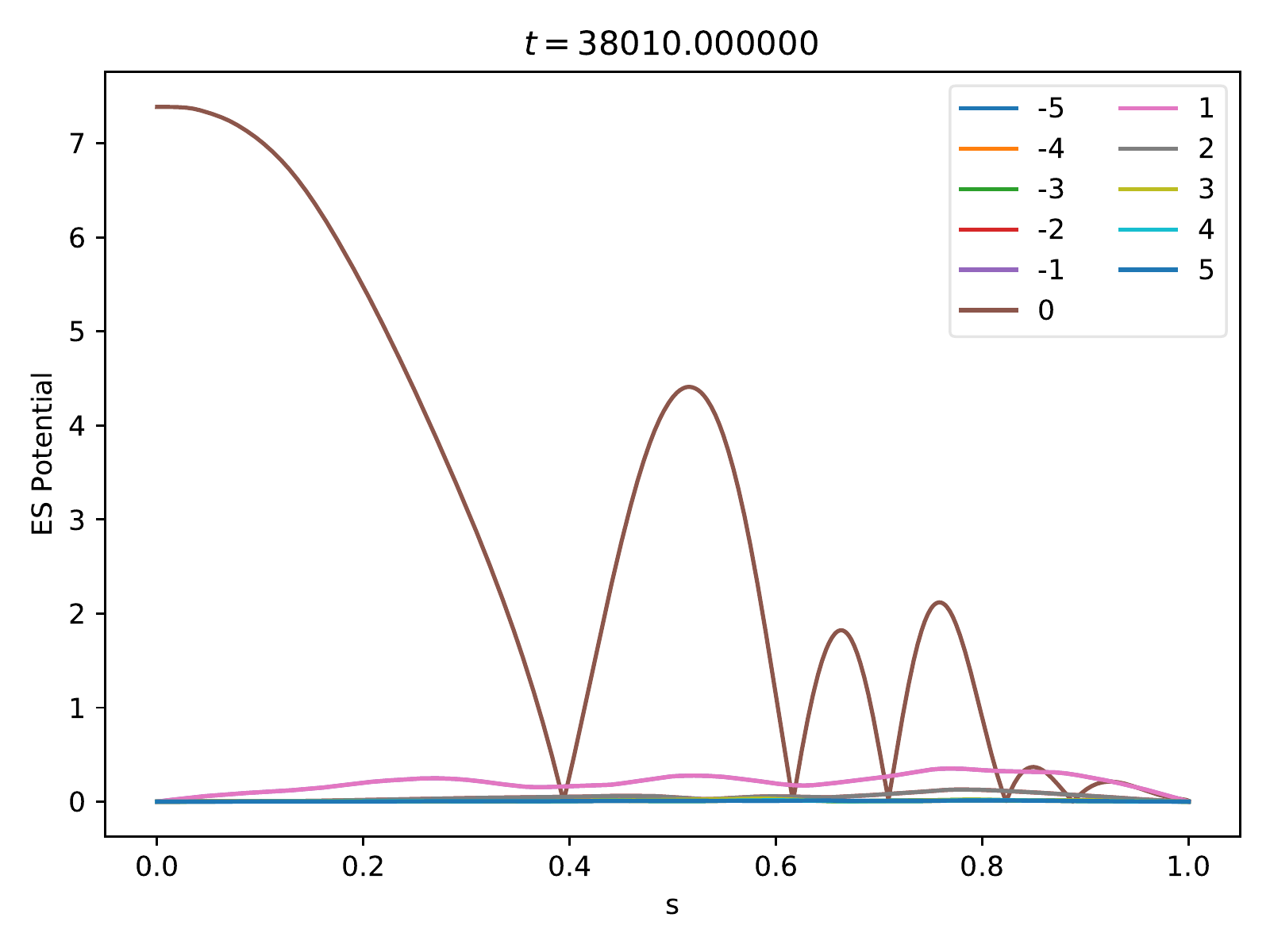}
{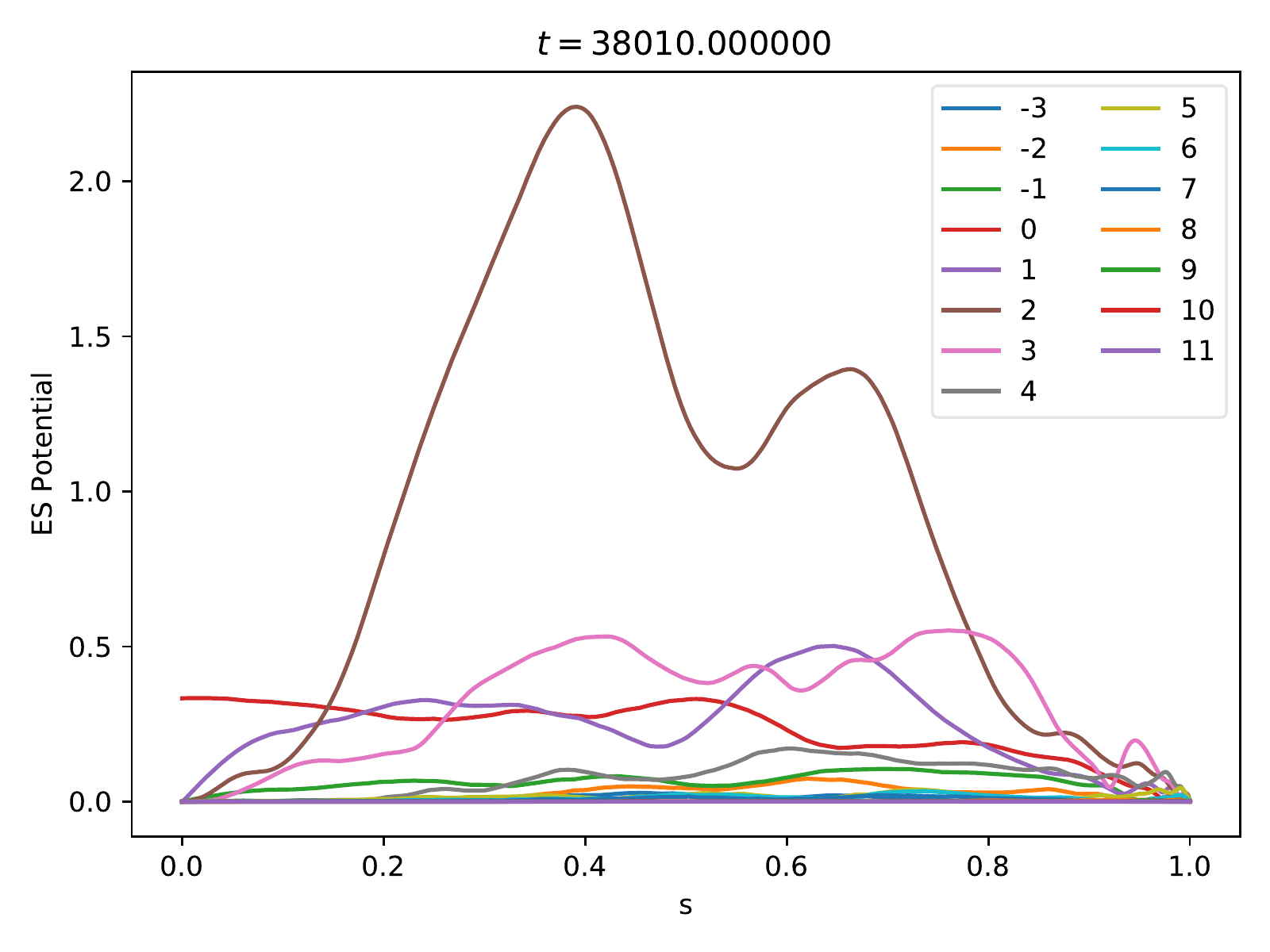}
{Radial dependence of the absolute value of $n=\{0\}$ 
(left) and $n=\{1\}$ (right) scalar potential components of the simulation with $n=\{0,1\}$ at low EPs concentration (see Fig.\ref{Fig:evolution_modes_ad_hoc} on the left). Note that the peaks of the $n=\{1\}$ ES potential correspond to the nodes of the $n=\{1\}$ ES potential, consistently with a model of the nonlinear forcing of the $n=\{1\}$ by the $n=\{0\}$. Color label correspond to different poloidal harmonics $m$.}
{Fig:mode_structure_ad_hoc}

\newpage

\subsection{Wave-wave interaction mediated by the EPs}\label{Sec:wave_wave-interaction}

The theoretical explanation is given in terms of a three-wave coupling of the ZS and AM, mediated by the curvature-pressure coupling term of the EPs. 
We refer to Ref.\cite{Qiu_2016}, where the study of ZFZF driven unstable by TAE was carried out. There, the most unstable mode (a TAE) was shown to trigger the ZFZF, through wave-wave coupling and the ZFZF grows with twice the growth rate of the TAE $\delta\phi_{ZFZF}\sim e^{2\gamma_{TAE}t}$.  Following the derivation there presented, we want to generalize it to a general AM driving a ZS and to investigate the inverse problem, that is the triggering of an AM driven unstable through wave-wave coupling by an EGAM (being in this case the most unstable mode: $\gamma^{L}_{EGAM}> \gamma^{L}_{AM}$), beating with the linear AM. 
To do so, we follow the theoretical derivation exposed in Ref.\cite{Qiu_2016}. However, here we start with the Vorticity equation for the AM (and not for the ZS), still neglecting the Reynolds and Maxwell stress, which express the non-linearity coming from the background plasma (in the present paper only the non-linearity in the EPs are retained): 
\begin{equation}
 \frac{c^{2}}{4\pi\omega_{AM}^{2}}B\frac{\partial}{\partial l}\frac{k^{2}_{\perp}}{B}\frac{\partial}{\partial l}\delta\psi_{AM}+\frac{e^{2}}{T_{i}}\langle (1-J_{k}^{2})F_{0}\rangle\delta\phi_{AM}  - \sum_{s}\langle \frac{e_{s}}{\omega}J_{k}\,\omega_{d}\,\delta H\rangle_{AM} = 0 
    \label{Eq:Vorticity}
\end{equation}

In Eq.\ref{Eq:Vorticity}, the first term represent the field line bending (FLB), the second term represents the Inertia Term (IT) and the third the Curvature Coupling Term (CCT). Here $e_{s}$ is the charge of the $s$-th species, $\partial/\partial_{l}$ is derivative parallel to the background magnetic field, $\delta H_{m,n}$ is the non-adiabatic response of the EPs associated to the scalar potential with polarization $(m,n)$. $\langle ...\rangle$ denotes velocity-space integration and $J_{k}=J_{0}(\gamma_{k})$ is the Bessel function of 0-order with argument $\gamma_{k}=k\rho$, being $k$ the mode wave-vector and $\rho$ the gyro-radius. In Eq.\ref{Eq:Vorticity}, the frequency drift $\omega_{d}$ appears. 
It satisfies:
\begin{equation}
\omega_{dZS}=\omega_{tr}\partial_{\theta}\lambda_{ZS}\quad \omega_{dAM}=\omega_{tr}\partial_{\theta}\lambda_{AM}
\label{Eq:omega_drift}
\end{equation}
where $\lambda_{dZS}$ and $\lambda_{dAM}$ define the coordinate transformation from the drift center to the particle gyro-center through the following relations: \begin{equation}
    \delta H^{NL}_{ZS}=e^{i\lambda_{dZS}}\delta H^{NL}_{dZS}\quad \delta H^{NL}_{AM}=e^{i\lambda_{dAM}}\delta H^{NL}_{dAM}\,.
\end{equation}
$\omega_{tr}=v_{\parallel}/(q\,R_{0})$ is the transit frequency. The non-adiabatic response of the EPs is calculated through the nonlinear gyrokinetic equation:
\begin{equation}
    (-i\omega+v_{\parallel}\partial_{l}+i\omega_{d})\delta H_{k} = -i\frac{e}{m}QF_{0}J_{k}\delta L_{k}-\frac{c}{B_{0}}\Lambda_{k}J_{k\prime}\delta L_{k\prime}\delta H_{k\prime\prime}
    \label{Eq:nonlinear_gyrokinetic}
\end{equation}

where $e$ is the charge of the EPs, the subscript $k$ is the wave-vector corresponding to the $(m,n)$ helicity, $\delta L = \delta \phi -\frac{v_{\parallel}}{c}\delta A_{\parallel}$ and $\Lambda_{\Vec{k}}=\sum_{\Vec{k}=\Vec{k'}+\Vec{k''}}\Vec{b}\cdot \Vec{k'}\cross\Vec{k''}$ is the coupling term. 
The first term on the right hand side of Eq.\ref{Eq:nonlinear_gyrokinetic} represents the linear response, while the second gives the nonlinear response. 
Characterizing the AM component of the CCT with its toroidal and polodial mode numbers $(m,n)$, 
we have that:
\begin{equation}
    CCT = \langle \frac{e}{\omega}J_{0}(\gamma_{AM})\omega_{d}\delta H^{NL}\rangle_{(m,n)}
    \label{Eq:CCT2}
\end{equation}
This can be rewritten as:
\begin{equation}
    CCT_{m,n}=\frac{e}{\omega}J_{0}(\gamma_{AM})e^{i(n\varphi-m\theta)}\langle \frac{1}{2  \pi}\int_{0}^{2\pi} d\varphi^{\prime}e^{-i\,n\varphi^{\prime}}\frac{1}{2  \pi}\int_{0}^{2\pi} d\theta^{\prime}e^{+i\,m\theta^{\prime}}\omega_{dAM}\delta H^{NL}_{m,n}\rangle
    \label{Eq:CCT3}
\end{equation}
Using Eq.\ref{Eq:omega_drift}, we can express Eq.\ref{Eq:CCT3} as:
\begin{equation}
    CCT_{m,n}=i\frac{e}{\omega}J_{0}(\gamma_{AM})e^{i(n\varphi-m\theta)}\langle \frac{1}{2  \pi}\int_{0}^{2\pi} d\varphi^{\prime}e^{-i\,n\varphi^{\prime}}\frac{1}{2  \pi}\int_{0}^{2\pi} d\theta^{\prime}e^{+i\,m\theta^{\prime}}e^{i\hat{\lambda}_{d n,m}\sin\theta^{\prime}}\partial_{\theta^{\prime}}\delta H^{NL}_{m,n}\rangle
\end{equation}
The nonlinear non-adiabatic response of the EPs to the mode $(m,n)$ is obtained from:
\begin{equation}
    (\partial_{t}+\omega_{tr}\partial_{\theta})\delta H^{NL}_{d,m,n}=-\frac{c}{B}e^{-i\lambda_{d}}\Lambda_{m,n}J_{0}(\gamma_{k\prime})\delta  L_{k\prime}\delta H_{k''} 
    \label{Eq:nonlinear_gyrokinetic_2}
\end{equation}

On the right hand side of Eq.\ref{Eq:nonlinear_gyrokinetic_2}, the term $\delta H_{k''}$ is the linear response of the EPs to the mode with wave-vector $k''$. This last couples two contributions of modes, expressed in $\delta  L_{k\prime}$ and $\delta H_{k''}$ and the following relations must be satisfied:
\begin{equation}
    n=n_{AM}+n_{EGAM}\quad ;\quad  m=m_{AM}+m_{EGAM}
\end{equation}
For the case  investigated in the present paper, we want to study how an AM $(2,1)$ is driven unstable by an EGAM, characterized by a scalar potential dominated by its $(0,0)$ component, but having $\delta H$ dominated by the mode numbers $(\pm 1,0)$. Note that the mechanism described here is valid in general for an AM with any helicity $(m_{AM},n_{AM})$.
Finally, with a similar calculation as that provided in Ref.\cite{Qiu_2016}, we obtain the following growth rate for the forced-driven AM:
\begin{equation}
    \delta \phi_{m,n}\sim e^{(\gamma_{AM}+\gamma_{EGAM})t}
    \label{Eq:sum_growth}
\end{equation}

Where $\gamma_{AM}$ and $\gamma_{EGAM}$ are the linear growth rates of the AM and of the EGAM, and the scalar potential on the left-hand-side of Eq.\ref{Eq:sum_growth}, refers to the nonlinearly generated AM, so:
\begin{equation}
    \gamma^{NL}_{AM}=\gamma_{AM}^{L}+\gamma_{EGAM}^{L}
\end{equation}
This explains how a marginally unstable Alfv\'en mode can be nonlinearly driven with a growth rate similar to that of the EGAM.


\newpage
\subsection{ASDEX Upgrade Equilibrium}\label{Sec:NLED-AUG_case}

In the present section, we show that the results obtained in the previous sections apply  also to a realistic ASDEX Upgrade magnetic equilibrium, AUG shot $\#$31213@0.84s. The plasma parameters and profiles are the same used in the previous sections, described in Sec.\ref{Sec:equilibrium}. 

\begin{figure}[!h]
     \centering
	\includegraphics[width=0.8\textwidth]{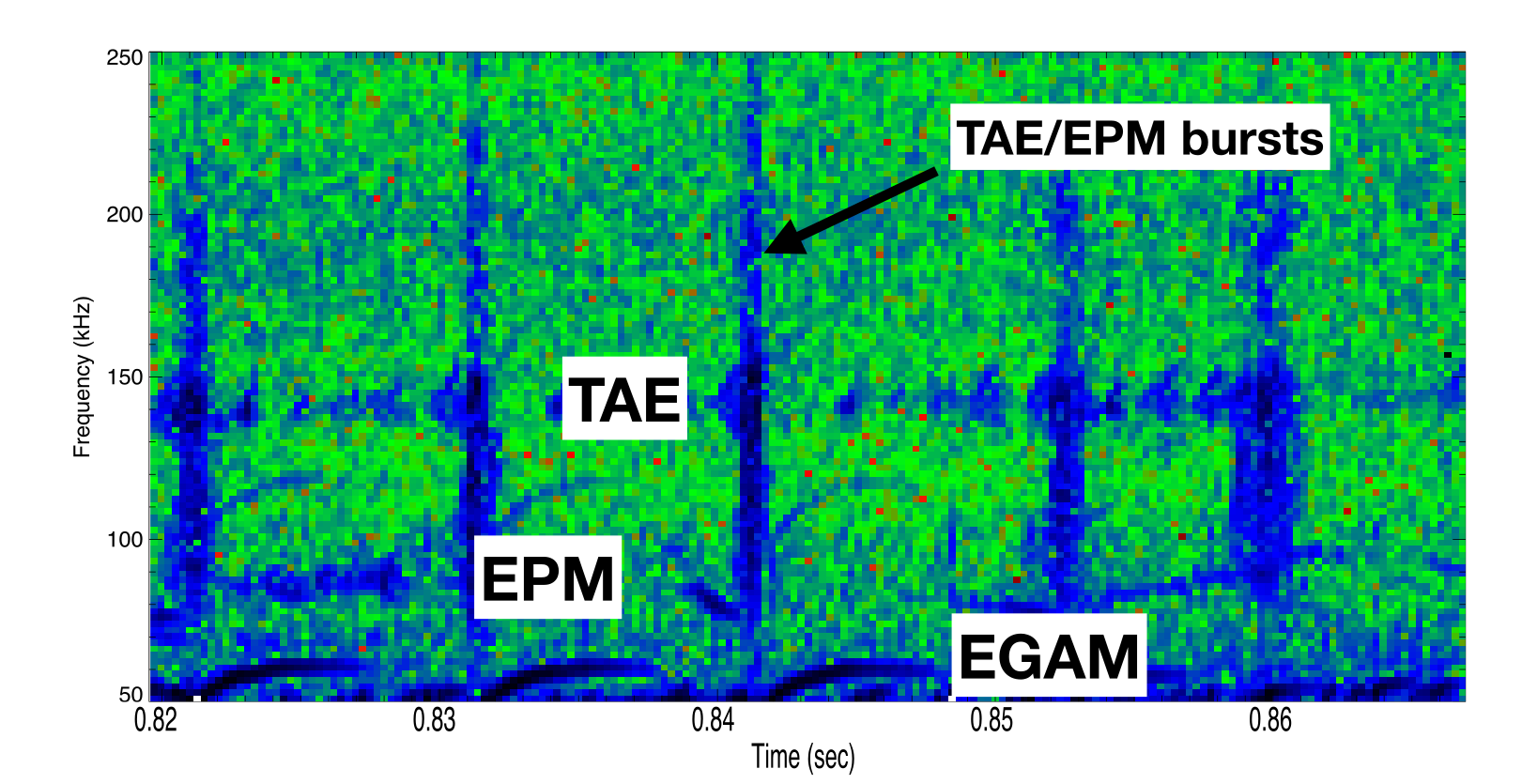}
	\caption{Detail of the experimental spectrogram obtained with Mirnov Coil. The magnetic equilibrium at $t=0.84\,s$ has been selected. }
	\label{Fig:spectrogram}
\end{figure}

This shot belongs to a series of discharges performed in 2017 in ASDEX Upgrade \cite{Lauber2}. The plasma scenario has been obtained through plasma parameters previously unexplored in ASDEX Upgrade: $\beta_{EP}/\beta_{th}\simeq 0.2-1$ and $\mathcal{E}_{EP}/T_{th}\simeq 150$ (the subscripts $th$ refers to the thermal plasma). Through these, a scenario where the mode dynamics is dominated by the EPs has been obtained and the turbulence contribution minimized. In these discharges the interaction between EGAMs and AMs has been observed. In order to choose the proper discharge, the beam injection angle dependence of the EGAM has been investigated. The discharge $\#31213$ has been selected to maximize the nonlinear EPs dynamics. This case, named the NLED-AUG case has become the base case for linear and non-linear EP simulations within three European theory projects \cite{Lauber}. The NLED-AUG case is obtained with early off-axis NB heating, injected with an angle between the horizontal axis and the beam-line of $7.13^{\circ}$. As it is shown in the portion of spectrogram in Fig.\ref{Fig:spectrogram}, an EPM-TAE burst is observed to trigger the EGAM. The experimental magnetic equilibrium measured at the time $t=0.84\,s$ is considered in the simulations of the present section. 

EPs have an  off-axis radial density profile (see Fig.\ref{Fig:profiles_radial}).
A scan against the EPs concentration  is presented, $5\% \leq \langle n_{EP}\rangle /\langle n_{D}\rangle\leq 10\%$. $\langle ..\rangle$ indicates the volume average. In this section the electrons have a realistic mass: $m_{e}\simeq m_{i}/3676$.

In Fig.\ref{Figure:nf_0053_0103} the time evolution of the scalar potential is presented, for simulations with low and high EPs
concentration. The dominant AM is an EPM, having the scalar potential dominant in its components $(m,n)=(2,1)$ and being mainly located around the radial position $s\simeq 0.2$ (as in Ref.\cite{VANNINI2020}). 

In the simulations where only a single toroidal mode is retained, the frequencies of the dominant modes in the linear phase are:
\begin{equation}
    \omega_{ZS}\simeq 0.055\,\omega_{A0} \quad \omega_{AM}\simeq 0.12\,\omega_{A0}
\end{equation}
corresponding respectively to $\nu_{ZS}\simeq 45 kHz$ and $\nu_{AM}\simeq 96 kHz$. The ZS is identified with an EGAM and it results to be in good agreement with what was detailed in Ref.\cite{NOVIKAU2020}. The frequency of the AM lies below the branch of the continuum spectrum $(m,n)=(2,1)$ and its frequency is in agreement with the experimental results, lying in the EPM-TAE burst. 

When both toroidal mode numbers are considered in the simulations, we observe the same tendency that was detailed in Sec.\ref{Sec:circular_equilibrium}. For high EPs concentration, we still observe the AM to drive the ZS, with a forced driven mechanism, before the saturation is reached. After that the first saturation of the AM is reached, the mode with the frequency of the EGAM is observed to develop.  Below $\langle n_{EP}\rangle/\langle n_{D}\rangle = 0.09$, instead, we observe the inverse process. An EGAM drives the AM, as it was discussed in Sec.\ref{Sec:wave_wave-interaction}.

In the regime below $\langle n_{EP}\rangle/\langle n_{D}\rangle = 0.09$, the difference of the saturation level of the AM with and without ZS is more pronounced (see Fig.\ref{Figure:Scan_nep}) with respect to higher concentrations of EPs. We note that in the former regime, the ZS are linearly more unstable than the AMs, whereas in the latter the AMs are more  unstable than the ZSs. The ZS in the former regime are identified as EGAMs.


\yFigTwo
{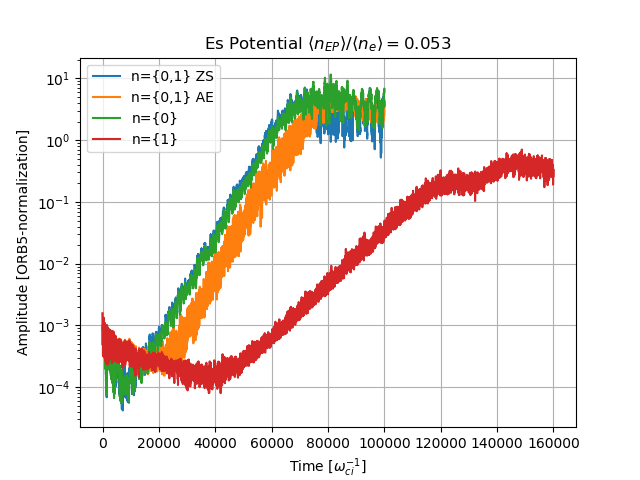}
{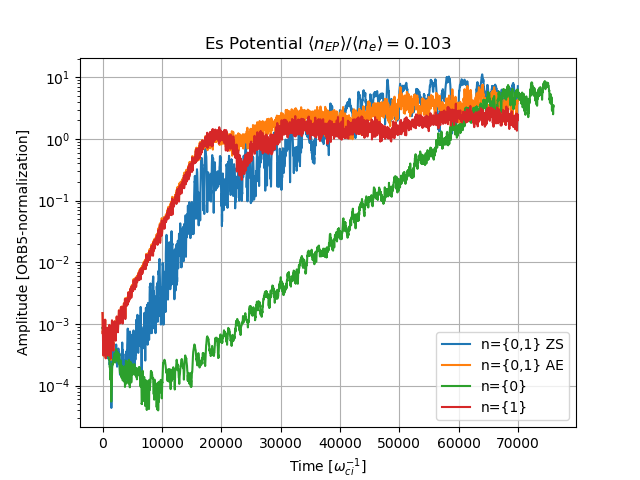}
{Modification of the AE dynamics in presence of the ZS, for low (left) and high (right) EPs concentration.}
{Figure:nf_0053_0103}




\yFigTwo
{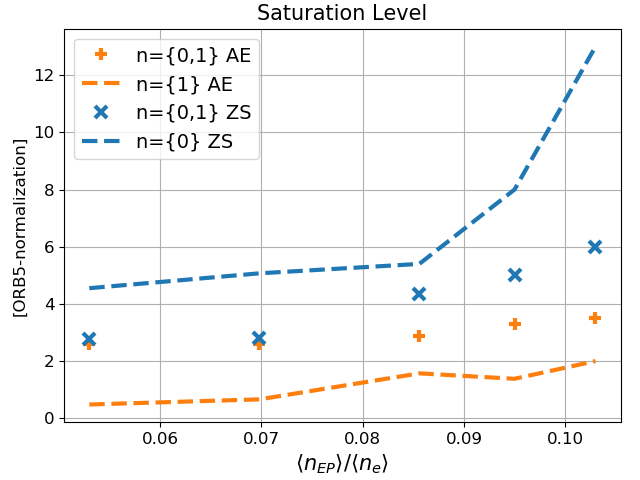}
{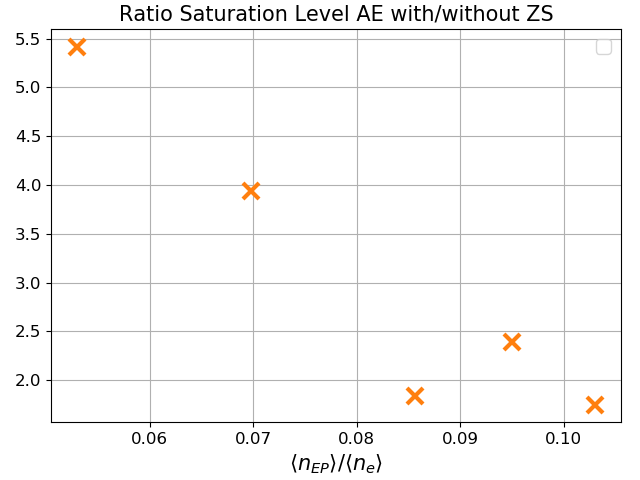}
{Modification of the saturation level of the AE, in presence of an EGAM. Note that strong modification of the 
AM saturation level in the regime of low EP concentrations, where the EGAM is stronger than the AM. On the right, the ratio of the saturation level of the AM in presence of the ZS against the corresponding value measured when the ZS is absent, is shown.}
{Figure:Scan_nep}

\clearpage

\newpage
\section{Conclusion}
The need to understand the nonlinear dynamics of EP-driven instabilities is motivated by the aim to be predictive about future scenarios that will be met in next generation fusion relevant machines. In this scenarios, an EP population is present due to external heating mechanisms or as a product of fusion reactions. Most of the previous theoretical works on the nonlinear dynamics of EPs driven instabilities has been focused on the independent evolution of single modes. The main saturation mechanisms considered in most of the previous studies was the EPs redistribution in real space or in phase space. 

In Ref.\cite{Todo_2010}, fully nonlinear magnetohydrodynamic simulations showed  that the saturation amplitude of a toroidal Alfv\'en eigenmode (TAE) is reduced by the non-linear generation of zonal modes. This was analytically explained in Ref.\cite{Qiu_2016},  where the formation of a zero frequency zonal structure (ZFZS) was found to be forced-driven by a pumping TAE which couples with its complex conjugate. The induced  EPs nonlinearities  was found to be dominant over those of the background plasma, which were expressed in the Maxwell and Reynolds stress of the vorticity equation. In both these studies the ZS were not excited directly by the EPs.




This paper has been dedicated to the study of the interaction between two different type of modes driven unstable by the EPs: EGAMs and AMs. The results of numerical simulations obtained with the nonlinear, gyrokinetic, electromagnetic, PIC code ORB5 have been presented. In the present work only the EPs have been allowed to follow their full trajectories, redistributing in the phase space. The nonlinearities arising from the background plasma have been here neglected. The EPs are, in this case, the main responsible of the saturation and mode interactions. Since the thermal plasma nonlinearities are not considered, the same arguments exposed in Ref.\cite{Qiu_2016} have been used to describe the interplay between the modes. The analytical derivation there exposed has been extended to the case here examined.

We observed that, in numerical simulations explained with analytical theory, through three wave coupling mediated by the EPs via the curvature pressure term in the vorticity equation, the nonlinear growth rate of the pumped mode is given by the sum of the linear growth rates of the pumping modes: $\gamma^{NL}_{n,m}=\gamma_{n',m'}^{L}+\gamma_{n'',m''}^{L}$.
EGAMs are found to channel energy into the AMs in the regime where the EGAMs linear growth rate is higher than the growth rate of the AM. Realistic ASDEX Upgrade plasma profiles have been considered. These results have been found both in analytical and experimental magnetic equilibrium. Despite the approximation of the EPs distribution function, these theoretical findings are qualitatively consistent with the experimental observations. This work opens the path to a new field of research, where the individual modes can saturate not only due to the EPs redistribution but also due to the mutual interaction.

In future works, simulations with more realistic distribution functions will be presented. Also all plasma nonlinearities will be taken into account, allowing to make comparisons with experimental data of ASDEX Upgrade, where an increased AM activity in the presence of strong EGAMs has also been observed experimentally.

\section{Acknowledgments}
Simulations presented in this work were performed on the CINECA Marconi supercomputer within the ORBFAST and OrbZONE projects. 

One of the authors, F. Vannini, would like to thank  Zhyiong Qiu for useful, interesting discussions and for great help provided in understanding the interaction between AM and ZS and for the help in deriving the exposed analytical theory. Also F. Vannini wants to thank Zhixin Lu and Omar Maj for the help provided in understanding aspects of the gyrokinetic theory. The authors wish to acknowledge stimulating discussions with F. Zonca, I. Novikau and A. Di Siena. 
This work was partly performed in the frame of the  \quotes{Multi-scale Energetic particle Transport in fusion devices} ER project.

This work has been carried out within the framework of the EUROfusion Consortium and has received funding from the Euratom research and training program 2014-2018 and 2019-2020 under grant agreement number 633053. The views and opinions expressed herein do not necessarily reflect those of the European Commission.

\clearpage
\bibliography{bibliography}
\bibliographystyle{unsrt}
\end{document}